\documentclass[english,jcp,preprint,endfloats,superscriptaddress]{revtex4-1} 
\usepackage{epsfig} 
\usepackage{amsmath} 
\usepackage{graphicx} 
\usepackage{longtable} 
\usepackage{hyperref} 
\usepackage{amssymb} 
\usepackage{dcolumn} 
\usepackage{mhchem} 
\usepackage{subfigure} 
\usepackage{array} 
\usepackage{multirow} 
\usepackage{tabularx} 
\usepackage{cleveref} 
\usepackage{mathrsfs}

\begin{document} 

\title{Short- and long-range corrected hybrid density functionals with the D3 dispersion corrections} 

\author{Chih-Wei Wang} 
\affiliation{Department of Physics, National Taiwan University, Taipei 10617, Taiwan} 

\author{Kerwin Hui} 
\affiliation{Department of Physics, National Taiwan University, Taipei 10617, Taiwan} 

\author{Jeng-Da Chai} 
\email[Author to whom correspondence should be addressed. Electronic mail: ]{jdchai@phys.ntu.edu.tw} 
\affiliation{Department of Physics, National Taiwan University, Taipei 10617, Taiwan} 
\affiliation{Center for Theoretical Sciences and Center for Quantum Science and Engineering, National Taiwan University, Taipei 10617, Taiwan} 

\date{\today} 

\begin{abstract} 

We propose a short- and long-range corrected (SLC) hybrid scheme employing 100\% Hartree-Fock (HF) exchange at both zero and infinite interelectronic distances, wherein three SLC hybrid density 
functionals with the D3 dispersion corrections (SLC-LDA-D3, SLC-PBE-D3, and SLC-B97-D3) are developed. SLC-PBE-D3 and SLC-B97-D3 are shown to be accurate for a very diverse range of 
applications, such as core ionization and excitation energies, thermochemistry, kinetics, noncovalent interactions, dissociation of symmetric radical cations, vertical ionization potentials, vertical 
electron affinities, fundamental gaps, and valence, Rydberg, and long-range charge-transfer excitation energies. Relative to $\omega$B97X-D, SLC-B97-D3 provides significant improvement for core 
ionization and excitation energies and noticeable improvement for the self-interaction, asymptote, energy-gap, and charge-transfer problems, while performing similarly for thermochemistry, kinetics, 
and noncovalent interactions. 

\end{abstract} 

\maketitle

\section{Introduction} 

Due to its decent balance between cost and performance, Kohn-Sham density functional theory (KS-DFT) \cite{HK,KS} has been a very popular electronic structure method for studying the ground-state 
properties of large systems \cite{Parr,DFTreview,DFTreview2,DFTreview3}. Recently, one of its most important extensions, time-dependent density functional theory (TDDFT) \cite{RG}, has also been 
actively developed for studying the excited-state and time-dependent properties of large systems \cite{Gross,Gross2,Casida,Casida2,TDDFT,TDDFT2,LR-TDDFT}. Nonetheless, the exact 
exchange-correlation (XC) energy functional $E_{xc}[\rho]$, which is the essential ingredient of both KS-DFT and adiabatic TDDFT, has not been found, and hence, density functional approximations 
(DFAs) for $E_{xc}[\rho]$ have been successively developed to improve the accuracy of KS-DFT and TDDFT for general applications. 

Functionals based on the conventional DFAs, such as the local density approximation (LDA) \cite{LDAX,LDAC}, generalized gradient approximations (GGAs) \cite{PBE}, and meta-GGAs 
(MGGAs) \cite{TPSS,SCAN}, are semilocal density functionals \cite{ladder}. They are reasonably accurate for the properties governed by short-range XC effects, and are computationally favorable for 
very large systems. Nevertheless, owing to the inadequate treatment of nonlocal XC effects \cite{Parr,DFTreview,DFTreview2,DFTreview3,SciYang}, semilocal density functionals can perform very 
poorly for the problems related to the self-interaction error (SIE) \cite{X2+}, noncovalent interaction error (NCIE) \cite{Dobson,D3review,SLR-vdW}, and static correlation error 
(SCE) \cite{MRreview,TAO1,TAO2,TAO3,TAO4,TAO5,TAO6}. 

In particular, some of these situations happen in the asymptotic regions ($r \to \infty$) of molecules, where the electron densities decay exponentially. In these regions, owing to the pronounced SIEs 
associated with semilocal density functionals, the functional derivatives of most semilocal density functionals (i.e., the semilocal XC potentials) do not exhibit the correct ($-1/r$) decay. Consequently, 
most semilocal density functionals can yield erroneous results for the highest occupied molecular orbital (HOMO) energies \cite{IP1,IP2,IP3,IP4,IP5,IP6} and high-lying Rydberg excitation 
energies \cite{Casida2,R10,R11,R12}. Even if the asymptote problems can be properly resolved by the recently developed semilocal density functionals with correct asymptotic 
behavior \cite{LFA,AK13,AC_Trickey,AC_Truhlar,p32} and asymptotically corrected model XC potentials \cite{LB94,LBa,Tozer,AA,AC1}, the SIE problems may remain unresolved \cite{p22}. Besides, 
semilocal density functionals are inaccurate for charge-transfer (CT) excitation energies \cite{p22,LFA,Peng,R20,R22new,new1,R12,R15,R16,R17,R18,R19}, due to the lack of a space- and 
frequency-dependent discontinuity in the adiabatic XC kernel adopted in TDDFT \cite{R22}. 

In 1993, on the basis of the adiabatic-connection formalism, Becke proposed global hybrid density functionals \cite{hybrid1,hybrid2}, combining semilocal density functionals with a small fraction 
(typically ranging from 0.2 to 0.25 for thermochemistry, and from 0.4 to 0.6 for kinetics) of Hartree-Fock (HF) exchange \cite{hybrid1,hybrid2,B3LYP,DFA0,B97,PBE0,PBE0a,BMK,M06-2X,SCAN0-2}. 
However, in certain situations, especially in the asymptotic regions of molecular systems, a very large fraction (even 100\%) of HF exchange is needed. Widely used global hybrid density functionals, 
such as B3LYP \cite{hybrid2,B3LYP}, PBE0 \cite{PBE0,PBE0a}, and M06-2X \cite{M06-2X}, do not qualitatively resolve the SIE, asymptote, and CT problems \cite{p22,p29}. 

With the aim of resolving these problems, long-range corrected (LC) hybrid density functionals \cite{LC,LCHirao,CAM-B3LYP,LC-wPBE,BNL,p6,p8,Herbert,wB97X-2,M11,p15,p19,wB97X-V,wB97M-V} 
have recently received considerable attention. A commonly used LC hybrid density functional (e.g., LC-$\omega$PBE \cite{LC-wPBE} and $\omega$B97 \cite{p6}) employs 100\% HF exchange for the 
long-range (LR) part of the interelectronic repulsion operator $\text{erf}(\omega r_{12})/r_{12}$, a semilocal exchange for the complementary short-range (SR) operator $\text{erfc}(\omega r_{12})/r_{12}$, 
and a semilocal correlation for the entire Coulomb operator $1/r_{12}$, with the parameter $\omega$ (typically ranging from 0.2 to 0.5 bohr$^{-1}$) specifying the partitioning of the interelectronic 
distance $r_{12} = |{\bf r}_{1} - {\bf r}_{2}|$ (atomic units are used throughout this paper). Here, erf is the standard error function, and erfc is the complementary error function. Besides, the inclusion of a 
small fraction of HF exchange at short range has been shown to improve the overall accuracy of conventional LC hybrid scheme (e.g., $\omega$B97X \cite{p6}). Over the years, LC hybrid functionals 
have been shown to qualitatively resolve the SIE, asymptote, and CT problems, offering a cost-effective way to incorporate nonlocal exchange effects. 

To properly account for noncovalent interactions, an accurate description of middle- and long-range dynamical correlation effects is essential. Accordingly, LC hybrid functionals can be combined with 
the DFT-D (KS-DFT with empirical dispersion corrections) schemes \cite{DFT-D1,DFT-D2,DFT-D3,DFT-D3ap,Sherrill,D3review} (e.g., $\omega$B97X-D \cite{p8}, $\omega$M05-D \cite{p15}, 
$\omega$M06-D3 \cite{p19}, and $\omega$B97X-D3 \cite{p19}) and the double-hybrid (adding a small fraction of second-order M\o ller-Plesset correlation) 
schemes \cite{B2PLYP,XYG3,DHrigo,PBE0-DH,TS2011,xDH-PBE0,PBE0-2,1DH-TPSS,PBE-ACDH,PBE-QIDH,QACF-2,CIDH,H-QIDH,SCAN0-2} (e.g., $\omega$B97X-2 \cite{wB97X-2}). Alternatively, 
LC hybrid functionals can also be incorporated with a fully nonlocal correlation density functional for van der Waals interactions (vdW-DF) \cite{vdW,VV10,VV10test} (e.g., 
$\omega$B97X-V \cite{wB97X-V} and $\omega$B97M-V \cite{wB97M-V}). Recently, we have shown that the $\omega$B97 series of functionals ($\omega$B97, $\omega$B97X, $\omega$B97X-D, etc.) 
has yielded impressive accuracy for various applications \cite{p22,p29,p25}, such as thermochemistry, kinetics, noncovalent interactions, dissociation of symmetric radical cations, frontier orbital energies, 
fundamental gaps, and valence, Rydberg, and long-range CT excitation energies. 

In spite of its general applicability, there are some situations, however, where the $\omega$B97 series can fail qualitatively. Very recently, Maier {\it et al.} \cite{lh} have shown that popular LC hybrid 
functionals, such as LC-$\omega$PBE and $\omega$B97X-D, perform very poorly for core excitation energies. They have also shown that global hybrid functionals with a large fraction (about 50\%) of 
HF exchange perform reasonably well for core excitation energies, showing consistency with the previous findings of Nakai and co-workers \cite{CV-B3LYP,CVR-B3LYP}. However, global hybrid 
functionals with 50\% HF exchange may not consistently perform well for thermochemistry and many other properties that do not require a large fraction of HF exchange. Within the framework of LC 
hybrid scheme, Hirao and co-workers have shown that the fraction of HF exchange at short range should be responsible for an accurate description of core excitation energies \cite{LCgau-coreBOP}. 
Similarly, the short-range corrected hybrid density functionals proposed by Besley {\it et al.} have been shown to accurately describe core excitation energies \cite{src}. 

On the other hand, Chai and Head-Gordon have shown that the fraction of HF exchange in the middle-range (MR) region (0.5 bohr $\lesssim r_{12} \lesssim$ 1.5 bohr) is important for a good balanced 
description of thermochemistry and kinetics \cite{p9}. Besides, they have argued that the fraction of HF exchange in the LR region ($r_{12} \gtrsim$ 1.5 bohr) should be crucial for the properties sensitive 
to the tail contributions (e.g., the SIE, asymptote, and CT problems), and the fraction of HF exchange in the SR region ($r_{12} \lesssim$ 0.5 bohr) should be responsible for the properties involving 
changes in the core contributions to $E_{xc}[\rho]$, such as core excitation energies. However, the SR region of the HF exchange operators adopted in the $\omega$B97 series has not been fully 
explored. Note that the fraction of HF exchange at zero interelectronic distance $r_{12} = 0$ is only 0.00, 0.16, 0.22, and 0.20 for $\omega$B97, $\omega$B97X, $\omega$B97X-D, and 
$\omega$B97X-D3, respectively. Nonetheless, as the electron densities in the core region are rather high (i.e., close to the high-density limit, where HF exchange should dominate correlation), we argue 
that a very large fraction of HF exchange in the SR region should be adopted for an accurate description of the properties sensitive to the core contributions (e.g., core ionization and excitation energies). 

In this work, we intend to improve the performance of the widely used LC hybrid functionals, LC-$\omega$PBE and the $\omega$B97 series, for core ionization and excitation energies, while retaining 
similar accuracy for many other applications. Specifically, we propose a new LC hybrid scheme employing 100\% HF exchange at $r_{12} = 0$ (i.e., the LC hybrid scheme is also short-range corrected), 
which is in strong contrast to the popular LC hybrid scheme (i.e., with the erf operator) and other LC hybrid schemes (e.g., with the erfgau \cite{OPfamily,Toulouse,LCgauBOP,LCgau-coreBOP} and 
terf \cite{terf,p11} operators) employing vanishing HF exchange at $r_{12} = 0$. The rest of this paper is organized as follows. 
We describe the short- and long-range corrected (SLC) hybrid scheme in Section II, and develop three SLC hybrid density functionals with the D3 dispersion corrections in Section III. 
The performance of our new functionals is compared with other functionals in Section IV (on the training set), and in Section V (on various test sets). Our conclusions are given in Section VI.

\section{Short- and Long-Range Corrected (SLC) Hybrid Scheme} 

In the SLC hybrid scheme, we first define the SLR operator $f_{\text{SLR}}(r_{12})/r_{12}$, which is an operator that approaches $1/r_{12}$ at both the SR ($r_{12} = 0$) and LR ($r_{12} \to \infty$) limits, 
and the complementary MR operator $f_{\text{MR}}(r_{12})/r_{12} = (1 - f_{\text{SLR}}(r_{12}))/r_{12}$ to partition the Coulomb operator: 
\begin{equation}\label{splitting} 
\frac{1}{r_{12}} = \frac{f_{\text{SLR}}(r_{12})}{r_{12}} + \frac{f_{\text{MR}}(r_{12})}{r_{12}}. 
\end{equation} 
In this work, we adopt 
\begin{equation}\label{fSLR} 
f_{\text{SLR}}(r_{12}) = \text{erfc}(\omega_{\text{SR}} r_{12}) + \text{erf}(\omega_{\text{LR}} r_{12}) 
\end{equation} 
as a simple sum of the SR function $\text{erfc}(\omega_{\text{SR}} r_{12})$ and LR function $\text{erf}(\omega_{\text{LR}} r_{12})$. Here, $\omega_{\text{SR}}$ and $\omega_{\text{LR}}$ are 
parameters controlling the SR and LR behavior, respectively, of $f_{\text{SLR}}(r_{12})$. Accordingly, we have 
\begin{equation}
\begin{split} 
f_{\text{MR}}(r_{12}) 
=&\; 1 - (\text{erfc}(\omega_{\text{SR}} r_{12}) + \text{erf}(\omega_{\text{LR}} r_{12})) \\ 
=&\; \text{erfc}(\omega_{\text{LR}} r_{12}) - \text{erfc}(\omega_{\text{SR}} r_{12}). 
\end{split} 
\end{equation} 

After the SLR/MR partition, a SLC hybrid density functional is defined as 
\begin{equation}\label{eq:SLC} 
E_{xc}^{\text{SLC}} = E_{x}^{\text{SLR-HF}} + E_{x}^{\text{MR-DFA}} + E_{c}^{\text{DFA}}. 
\end{equation} 
Here, 
$E_{c}^{\text{DFA}}$ is the DFA correlation energy of the Coulomb operator $1/r_{12}$, $E_{x}^{\text{SLR-HF}}$ is the HF exchange energy of the SLR operator 
$f_{\text{SLR}}(r_{12})/r_{12} = \text{erfc}(\omega_{\text{SR}} r_{12})/r_{12} + \text{erf}(\omega_{\text{LR}} r_{12})/r_{12}$ (computed by the occupied Kohn-Sham (KS) orbitals $\{\psi_{i\sigma}({\bf r})\}$): 
\begin{equation}\label{eq:SLR-HF} 
\begin{split} 
E_{x}^{\text{SLR-HF}} 
=&\; - \frac{1}{2} \sum_{\sigma}^{\alpha,\beta} \sum_{i,j}^{occ.} \iint 
\psi_{i\sigma}^{*}({\bf r}_{1}) \psi_{j\sigma}^{*}({\bf r}_{2}) \frac{f_{\text{SLR}}(r_{12})}{r_{12}} \psi_{j\sigma}({\bf r}_{1}) \psi_{i\sigma}({\bf r}_{2}) d{\bf r}_{1} d{\bf r}_{2} \\ 
=&\; E_{x}^{\text{SR-HF}}(\omega_{\text{SR}}) + E_{x}^{\text{LR-HF}}(\omega_{\text{LR}}), 
\end{split} 
\end{equation} 
where 
\begin{equation}\label{eq:SR-HF} 
E_{x}^{\text{SR-HF}}(\omega_{\text{SR}}) = - \frac{1}{2} \sum_{\sigma}^{\alpha,\beta} \sum_{i,j}^{occ.} \iint 
\psi_{i\sigma}^{*}({\bf r}_{1}) \psi_{j\sigma}^{*}({\bf r}_{2}) \frac{\text{erfc}(\omega_{\text{SR}} r_{12})}{r_{12}} \psi_{j\sigma}({\bf r}_{1}) \psi_{i\sigma}({\bf r}_{2}) d{\bf r}_{1} d{\bf r}_{2} 
\end{equation} 
is the HF exchange energy of the SR operator $\text{erfc}(\omega_{\text{SR}} r_{12})/r_{12}$, and 
\begin{equation}\label{eq:LR-HF} 
E_{x}^{\text{LR-HF}}(\omega_{\text{LR}}) = - \frac{1}{2} \sum_{\sigma}^{\alpha,\beta} \sum_{i,j}^{occ.} \iint 
\psi_{i\sigma}^{*}({\bf r}_{1}) \psi_{j\sigma}^{*}({\bf r}_{2}) \frac{\text{erf}(\omega_{\text{LR}} r_{12})}{r_{12}} \psi_{j\sigma}({\bf r}_{1}) \psi_{i\sigma}({\bf r}_{2}) d{\bf r}_{1} d{\bf r}_{2} 
\end{equation} 
is the HF exchange energy of the LR operator $\text{erf}(\omega_{\text{LR}} r_{12})/r_{12}$. In addition, $E_{x}^{\text{MR-DFA}}$, the DFA exchange energy of 
the MR operator $f_{\text{MR}}(r_{12})/r_{12} = \text{erfc}(\omega_{\text{LR}} r_{12})/r_{12} - \text{erfc}(\omega_{\text{SR}} r_{12})/r_{12}$, can be expressed as 
\begin{equation}\label{eq:MR-DFA} 
E_{x}^{\text{MR-DFA}} = E_{x}^{\text{SR-DFA}}(\omega_{\text{LR}}) - E_{x}^{\text{SR-DFA}}(\omega_{\text{SR}}), 
\end{equation} 
where $E_{x}^{\text{SR-DFA}}(\omega)$ is the DFA exchange energy of the SR operator $\text{erfc}(\omega r_{12})/r_{12}$. 

On the basis of Eq.\ (\ref{eq:SLR-HF}), $f_{\text{SLR}}(r_{12})$ can be regarded as the fraction of HF exchange at $r_{12}$ for the SLC hybrid density functional. Therefore, we impose the constraint 
$0 \le \omega_{\text{LR}} \le \omega_{\text{SR}} < \infty$ to ensure that $0 \le f_{\text{MR}}(r_{12}) \le 1$ and hence, $0 \le f_{\text{SLR}}(r_{12}) \le 1$ can be satisfied at each $r_{12}$. 
For $\omega_{\text{SR}} = \omega_{\text{LR}}$, we have $f_{\text{MR}}(r_{12}) = 0$ and $f_{\text{SLR}}(r_{12}) = 1$, employing the full HF exchange and a DFA correlation. 
Note that $f_{\text{SLR}}(r_{12})$ (given by Eq.\ ({\ref{fSLR})) provides a smooth transition between the following two limits: 
\begin{equation} 
f_{\text{SLR}}(r_{12} = 0) = 1,\ 
\lim_{r_{12} \to \infty} f_{\text{SLR}}(r_{12}) = 1, 
\end{equation} 
employing 100\% HF exchange at both the SR ($r_{12} = 0$) and LR ($r_{12} \to \infty$) limits. 
Note also that the SLC hybrid scheme reduces to the popular LC hybrid scheme (i.e., with the erf operator) as $\omega_{\text{SR}} \to \infty$, while 
it reduces to pure KS-DFT as $\omega_{\text{SR}} \to \infty$ and $\omega_{\text{LR}} = 0$.

\section{SLC Hybrid Functionals with Dispersion Corrections} 

On the basis of Eq.\ (\ref{eq:SLC}), here we introduce three SLC hybrid density functionals with the D3 dispersion corrections. 
As the simplest DFA is the LDA, we define the SLC-LDA functional as 
\begin{equation}\label{eq:slc-lda} 
E_{xc}^{\text{SLC-LDA}} = E_{x}^{\text{SLR-HF}} + E_{x}^{\text{MR-LDA}} + E_{c}^{\text{LDA}}, 
\end{equation} 
where 
$E_{c}^{\text{LDA}}$ is the LDA correlation functional \cite{LDAC}, 
$E_{x}^{\text{SLR-HF}}$ is the SLR-HF exchange energy (given by Eq.\ (\ref{eq:SLR-HF})), and 
\begin{equation}\label{eq:MR-LDA} 
E_{x}^{\text{MR-LDA}} 
= E_{x}^{\text{SR-LDA}}(\omega_{\text{LR}}) - E_{x}^{\text{SR-LDA}}(\omega_{\text{SR}}) 
= \sum_{\sigma}^{\alpha,\beta} \int e_{x\sigma}^\text{MR-LDA} d{\bf r} 
\end{equation} 
is the MR-LDA exchange functional, which is known due to the analytical form of 
$E_x^{\text{SR-LDA}}(\omega)$, the LDA exchange functional of the SR operator $\text{erfc}(\omega r_{12})/r_{12}$ \cite{LC,Gill96}. 
Here, $e_{x\sigma}^\text{MR-LDA}$ is the MR-LDA exchange energy density for $\sigma$-spin, 
\begin{equation}\label{eq:MR-LDAD} 
e_{x\sigma}^\text{MR-LDA} = -\frac{3}{2}\left(\frac{3}{4\pi}\right)^{1/3}\rho_\sigma^{4/3}({\bf r})\left[F(a_{\text{LR},\sigma}) - F(a_{\text{SR},\sigma})\right], 
\end{equation} 
where $a_{\text{LR},\sigma} \equiv \omega_{\text{LR}}/(2(6\pi^2\rho_\sigma({\bf r}))^{1/3})$ and $a_{\text{SR},\sigma} \equiv \omega_{\text{SR}}/(2(6\pi^2\rho_\sigma({\bf r}))^{1/3})$ 
are dimensionless parameters controlling the values of the attenuation function $F(a)$, 
\begin{equation} 
F(a) = 1 - \frac{8}{3} a \left[\sqrt{\pi} \text{erf}\left(\frac{1}{2a}\right) - 3a + 4a^3 + (2a - 4a^3) \text{exp}\left(-\frac{1}{4a^2}\right)\right]. 
\end{equation} 

To go beyond the simplest SLC-LDA, we define the SLC-PBE functional as 
\begin{equation}\label{eq:slc-pbe} 
E_{xc}^{\text{SLC-PBE}} = E_{x}^{\text{SLR-HF}} + E_{x}^{\text{MR-PBE}} + E_{c}^{\text{PBE}}, 
\end{equation} 
where 
$E_{c}^{\text{PBE}}$ is the PBE correlation functional \cite{PBE}, 
$E_{x}^{\text{SLR-HF}}$ is the SLR-HF exchange energy (given by Eq.\ (\ref{eq:SLR-HF})), and 
\begin{equation}\label{eq:MR-PBE} 
E_{x}^{\text{MR-PBE}} = E_{x}^{\text{SR-PBE}}(\omega_{\text{LR}}) - E_{x}^{\text{SR-PBE}}(\omega_{\text{SR}}) 
\end{equation} 
is the MR-PBE exchange functional, with $E_{x}^{\text{SR-PBE}}(\omega)$ being the PBE exchange functional of the SR operator $\text{erfc}(\omega r_{12})/r_{12}$ \cite{HJS}. 

To further improve upon SLC-PBE, we adopt flexible functional forms in Eq.\ (\ref{eq:SLC}). 
Similar to the B97 ansatz \cite{B97}, we define the SLC-B97 functional as 
\begin{equation}\label{eq:slc-b97} 
E_{xc}^{\text{SLC-B97}} = E_{x}^{\text{SLR-HF}} + E_{x}^{\text{MR-B97}} + E_{c}^{\text{B97}}. 
\end{equation} 
Here, $E_{c}^{\text{B97}}$ has the same functional form as the B97 correlation functional \cite{B97}, which can be decomposed into 
same-spin $E_{c\sigma\sigma}^{B97}$ and opposite-spin $E_{c\alpha\beta}^{B97}$ components, 
\begin{equation} 
E_{c}^{\text{B97}} = \sum_{\sigma}^{\alpha,\beta} E_{c\sigma\sigma}^{\text{B97}} + E_{c\alpha\beta}^{\text{B97}}. 
\end{equation} 
Here, 
\begin{equation}\label{eq:b97css} 
E_{c\sigma\sigma}^{\text{B97}} = \int 
e_{c\sigma\sigma}^{\text{LDA}}\ \sum_{i=0}^{m} c_{c\sigma\sigma,i} \left(\frac{\gamma_{c\sigma\sigma} s_{\sigma}^2}{1+\gamma_{c\sigma\sigma} s_{\sigma}^2}\right)^{i} d{\bf r}, 
\end{equation} 
\begin{equation}\label{eq:b97cab} 
E_{c\alpha\beta}^{\text{B97}} = \int 
e_{c\alpha\beta}^{\text{LDA}}\ \sum_{i=0}^{m} c_{c\alpha\beta,i} \left(\frac{\gamma_{c\alpha\beta} s_{av}^2}{1+\gamma_{c\alpha\beta} s_{av}^2}\right)^{i} d{\bf r}, 
\end{equation} 
where $\gamma_{c\sigma\sigma} = 0.2$, $\gamma_{c\alpha\beta} = 0.006$, $s_{av}^{2} = \frac{1}{2}(s_{\alpha}^{2}+s_{\beta}^{2})$, and 
$s_{\sigma} = |\nabla \rho_{\sigma}({\bf r})|/\rho_{\sigma}^{4/3}({\bf r})$. The correlation energy densities $e_{c\sigma\sigma}^{\text{LDA}} = e_{c}^{\text{LDA}}(\rho_{\sigma},0)$ and 
$e_{c\alpha\beta}^{\text{LDA}} = e_{c}^{\text{LDA}}(\rho_{\alpha},\rho_{\beta}) - e_{c}^{\text{LDA}}(\rho_{\alpha},0) - e_{c}^{\text{LDA}}(0,\rho_{\beta})$ are derived from the PW92 parametrization of 
the LDA correlation energy density $e_{c}^{\text{LDA}}(\rho_{\alpha},\rho_{\beta})$ \cite{LDAC}, using the approach of Stoll {\it et al.} \cite{Stoll2}. 
In addition, $E_{x}^{\text{SLR-HF}}$ is the SLR-HF exchange energy (given by Eq.\ (\ref{eq:SLR-HF})), and 
\begin{equation}\label{eq:MR-B97} 
E_{x}^{\text{MR-B97}} = \sum_{\sigma}^{\alpha,\beta} \int 
e_{x\sigma}^\text{MR-LDA}\ \sum_{i=0}^{m} c_{x \sigma,i} \left(\frac{\gamma_{x\sigma} s_{\sigma}^2}{1+\gamma_{x\sigma} s_{\sigma}^2}\right)^{i} d{\bf r} 
\end{equation} 
is the MR-B97 exchange functional, where $\gamma_{x\sigma} = 0.004$ and $e_{x\sigma}^\text{MR-LDA}$ is given by Eq.\ (\ref{eq:MR-LDAD}). Note that $E_{x}^{\text{MR-B97}}$ 
has the same functional form as the SR-B97 exchange functional (see Eq.\ (11) of Ref.\ \cite{p6}) when $\omega_{\text{SR}} \to \infty$, and 
has the same functional form as the B97 exchange functional \cite{B97} when $\omega_{\text{SR}} \to \infty$ and $\omega_{\text{LR}} = 0$. 

Following the DFT-D3 scheme \cite{DFT-D3}, our total energy is given by 
\begin{equation}\label{eq:dft-d3} 
E_{\text{DFT-D3}} = E_{\text{KS-DFT}} + E_{\text{disp}}(\text{D3}), 
\end{equation} 
where $E_{\text{KS-DFT}}$ is the total energy in KS-DFT, and 
\begin{equation}\label{eq:d3} 
E_{\text{disp}}(\text{D3}) = - \sum_{n=6,8} \sum_{A > B} \frac{C_n^{AB}}{R_{AB}^{n}[1+6(s_{r,n} R_{0}^{AB}/R_{AB})^{n+8}]} 
\end{equation} 
is the D3 dispersion correction (the unscaled version is adopted, and the three-body term is not included). Here, the second sum is over all atom pairs in the system, and $R_{AB}$ is the interatomic 
distance of atom pair $AB$, while the cutoff radius $R_0^{AB}$ and the dispersion coefficients ($C_6^{AB}$ and $C_8^{AB}$) for atom pair $AB$ are provided in the DFT-D3 scheme \cite{DFT-D3}. 
Therefore, $s_{r,6}$ and $s_{r,8}$, which control the strength of dispersion correction, are the parameters to be determined. 

In this work, the SLC-LDA (Eq.\ (\ref{eq:slc-lda})), SLC-PBE (Eq.\ (\ref{eq:slc-pbe})), and SLC-B97 (Eq.\ (\ref{eq:slc-b97})) functionals with the D3 dispersion corrections (Eq.\ (\ref{eq:d3})) are denoted 
as SLC-LDA-D3, SLC-PBE-D3, and SLC-B97-D3, respectively. Note that SLC-LDA-D3 and SLC-PBE-D3 satisfy the exact uniform electron gas (UEG) limit by construction, while the exact UEG limit 
for SLC-B97-D3 is enforced by imposing the following constraints: $c_{x\sigma,0} = c_{c\sigma\sigma,0} = c_{c\alpha\beta,0} = 1$. 

The four parameters ($\omega_{\text{SR}}$, $\omega_{\text{LR}}$, $s_{r,6}$, and $s_{r,8}$) of SLC-LDA-D3 and SLC-PBE-D3 are determined by least-squares fittings to 
the accurate experimental and theoretical data in the training set, involving 
\begin{itemize} 
\item the 223 atomization energies (AEs) of the G3/99 set \cite{G399}, 
\item the 40 ionization potentials (IPs), 25 electron affinities (EAs), and 8 proton affinities (PAs) of the G2-1 set \cite{G21}, 
\item the 76 barrier heights of the NHTBH38/04 and HTBH38/04 sets \cite{BH}, 
\item the 22 noncovalent interactions of the S22 set \cite{S22,S22b}. 
\end{itemize} 
For the S22 set, an updated version of reference values from S22B \cite{S22b} are adopted. For the parameter optimization, we focus on a range of possible 
$\omega_{\text{SR}}$ (0.8, 1.0, 1.5, 2.0, 2.5, 3.0, 3.5, and 4.0 bohr$^{-1}$) and $\omega_{\text{LR}}$ (0.30, 0.35, 0.40, 0.45, and 0.50 bohr$^{-1}$) values, and optimize the corresponding 
$s_{r,6}$ and $s_{r,8}$ in steps of 0.001, for $0 < s_{r,6} < 2$ and $0 < s_{r,8} < 2$, respectively. The S22 data are weighted 10 times more than the others. 
As is usual in hybrid density functional approaches, the electronic energy is minimized with respect to the orbitals. Detailed information about the training set can be found in Refs.\ \cite{p6,p15,p19}. 

The optimized parameters of SLC-LDA-D3 and SLC-PBE-D3 are summarized in \Cref{table:para}, and the HF exchange operators adopted in SLC-LDA-D3, SLC-PBE-D3, and the $\omega$B97 
series are plotted in \Cref{fig:erf}. Note that the HF exchange operators adopted in LC-$\omega$PBE and $\omega$B97 are the same. As can be seen, the fractions of HF exchange adopted in 
SLC-PBE-D3 and the $\omega$B97 series are similar in the MR region, showing consistency with the previous findings of Chai and Head-Gordon \cite{p9} that the fine details of the MR region of 
the HF exchange operators adopted are important for good balanced performance in thermochemistry and kinetics. Besides, as the LR-HF exchange contributions (see Eq.\ (\ref{eq:LR-HF})) in 
SLC-PBE-D3, LC-$\omega$PBE, and $\omega$B97 are the same (with $\omega_{\text{LR}}$ = 0.40 $\text{bohr}^{-1}$), SLC-PBE-D3, LC-$\omega$PBE, and $\omega$B97 should have similar 
performance for the properties sensitive to the tail contributions. In addition, the SR-HF exchange contribution (see Eq.\ (\ref{eq:SR-HF})) in SLC-PBE-D3 is significant only in the region of 
$r_{12} \lesssim 1/\omega_{\text{SR}}$ = 0.5 bohr (i.e., the same as the SR region identified by Chai and Head-Gordon \cite{p9}), and hence, should be responsible only for the properties sensitive 
to the core contributions. By contrast, for SLC-LDA-D3, a larger fraction of HF exchange is needed to reduce the severe error associated with the underlying LDA. Interestingly, the HF exchange 
operators adopted in SLC-LDA-D3 and SLC-PBE-D3 look upside down, when compared with those adopted in the MR hybrid functionals developed by Henderson {\it et al.} for different 
purposes \cite{HISS}. 

As the HF exchange operator adopted in SLC-PBE-D3 has been optimized, the same HF exchange operator is adopted in SLC-B97-D3 without further optimization. However, the remaining D3 
parameters ($s_{r,6}$ and $s_{r,8}$) and B97 linear expansion coefficients ($c_{x\sigma,i}$, $c_{c\sigma\sigma,i}$, and $c_{c\alpha\beta,i}$) of SLC-B97-D3 are determined self-consistently by 
a least-squares fitting procedure described in Ref.\ \cite{p6} (using the same training set), with the SLC-PBE-D3 orbitals being the initial guess orbitals. During the parameter optimization, 
as the statistical errors of the training set for SLC-B97-D3 are not significantly improved for $m > 4$, the functional expansions adopted in SLC-B97-D3 are truncated at $m = 4$. 
We summarize the optimized parameters of SLC-B97-D3 in \Cref{table:para}. 

In the following sections, the overall performance of SLC-LDA-D3, SLC-PBE-D3, and SLC-B97-D3 will be compared with a popular semilocal functional: 
\begin{itemize} 
\item PBE \cite{PBE}, 
\end{itemize} 
and several widely used LC hybrid functionals: 
\begin{itemize} 
\item LC-$\omega$PBE \cite{LC-wPBE}, 
\item $\omega$B97 \cite{p6}, 
\item $\omega$B97X \cite{p6}, 
\item $\omega$B97X-D \cite{p8}, 
\item $\omega$B97X-D3 \cite{p19} 
\end{itemize} 
on the training set and various test sets \cite{supp}.

\section{Results for the Training Set} 

All calculations are performed with a development version of \textsf{Q-Chem 4.3} \cite{QChem}. Spin-restricted theory is used for singlet states and spin-unrestricted theory for others, unless noted 
otherwise. For the interaction energies of the weakly bound systems, the counterpoise correction \cite{CP} is employed to reduce the basis set superposition error (BSSE). 

Results for the training set are computed using the 6-311++G(3df,3pd) basis set with the fine grid EML(75,302), consisting of 75 Euler-Maclaurin radial grid points \cite{EM} and 302 Lebedev angular 
grid points \cite{L}. The error for each entry is defined as error = theoretical value $-$ reference value. The notation adopted for characterizing statistical errors is as follows: 
mean signed errors (MSEs), mean absolute errors (MAEs), and root-mean-square (rms) errors. 

As shown in \Cref{table:train}, SLC-PBE-D3 consistently outperforms PBE and LC-$\omega$PBE for the AEs of the G3/99 set and noncovalent interactions of the S22 set, reflecting the effect of the 
improved HF exchange operator and dispersion correction, respectively. 
To provide the fairest comparison to SLC-PBE-D3, the performance of LC-$\omega$PBE-D3 (i.e., LC-$\omega$PBE with the D3 dispersion correction) \cite{DFT-D3ap} is also examined here. 
While LC-$\omega$PBE-D3 performs similarly to SLC-PBE-D3 for the S22 set due to the inclusion of dispersion correction, LC-$\omega$PBE-D3 performs considerably worse than SLC-PBE-D3 
for the G3/99 set. 

Owing to its flexible functional forms, SLC-B97-D3 generally outperforms SLC-PBE-D3, and significantly outperforms SLC-LDA-D3 on the training set. 
Besides, as the fractions of HF exchange adopted in SLC-B97-D3 and $\omega$B97 are similar in the MR region, SLC-B97-D3 performs similarly to $\omega$B97 for thermochemistry and kinetics, 
implying that the SR-HF exchange contribution in SLC-B97-D3 does not degrade its performance for normal chemistry. 
However, as mentioned previously, the HF exchange operator of SLC-PBE-D3 is adopted in SLC-B97-D3 (i.e., without further optimization), though the D3 parameters and B97 linear expansion 
coefficients of SLC-B97-D3 are optimized on the training set. Therefore, SLC-B97-D3 performs slightly worse than $\omega$B97X-D3 (where the HF exchange operator, D3 parameters, and B97 
linear expansion coefficients were fully optimized on the same training set). All the dispersion-corrected functionals perform reasonably well for the noncovalent interactions of the S22 set.

\section{Results for the Test Sets} 

To examine how SLC-LDA-D3, SLC-PBE-D3, and SLC-B97-D3 perform outside the training set, we also assess their performance on a wide variety of test sets, including 
\begin{itemize} 
\item the 23 core ionization energies of 14 molecules \cite{CI}, 
\item the 38 core excitation energies of 13 molecules \cite{src}, 
\item the 66 noncovalent interactions of the S66 set \cite{S66}, 
\item four dissociation energy curves of symmetric radical cations \cite{X2+}, 
\item the 113 AEs of the AE113 database \cite{p15,p22}, 
\item the 131 vertical IPs of the IP131 database \cite{p15}, 
\item the 131 vertical EAs of the EA131 database \cite{p15,p22}, 
\item the 131 fundamental gaps of the FG131 database \cite{p15,p22}, 
\item the 19 valence and 23 Rydberg excitation energies of five molecules \cite{ex}, 
\item one long-range CT excitation energy curve of two well-separated molecules \cite{R15,CT_ref}. 
\end{itemize} 
As will be discussed later, each vertical IP can be computed in two different ways, each vertical EA can be computed in three different ways, and each fundamental gap can be computed 
in three different ways. Consequently, there are in total 1335 pieces of data in the test sets, which are larger and more diverse than the training set.

\subsection{Core ionization energies} 

To assess the accuracy of the density functionals on core ionization energies, the 23 core ionization energies of 14 molecules are collected from Ref.\ \cite{CI}, where the atoms at which the $1s$ 
electrons are ionized are all first-row elements. As discussed by Baerends and co-workers \cite{Baerends}, the ionization energies for all the occupied orbitals can be well approximated by 
the minus orbital energies, when the exact (or highly accurate) XC potential is adopted. Therefore, in this work, the core ionization energy of a molecule is calculated as the minus core orbital energy 
of the molecule, using the 6-311++G(3df,3pd) basis set and EML(75,302) grid. 

As shown in \Cref{table:ci}, PBE performs worst for the core ionization energies, while LC-$\omega$PBE and $\omega$B97 only have minor improvement due to the vanishingly small fraction of HF 
exchange at small interelectronic distances. Besides, $\omega$B97X, $\omega$B97X-D, and $\omega$B97X-D3, which include a small fraction of SR-HF exchange, perform slightly better than 
LC-$\omega$PBE and $\omega$B97. Among the functionals examined on the core ionization energies, SLC-B97-D3 ranks first, while SLC-PBE-D3 and SLC-LDA-D3 rank second and third, 
respectively. Overall, the SLC hybrid functionals are comparable in performance, and are much more accurate than PBE, LC-$\omega$PBE, the $\omega$B97 series, and possibly, other LC hybrid 
functionals employing a small fraction of HF exchange in the SR region, reflecting that a very large fraction of HF exchange in the SR region is indeed essential for an accurate description of core 
ionization energies. While the relativistic corrections are not considered here, our comments remain the same for the core ionization energies with the relativistic corrections \cite{supp}.

\subsection{Core excitation energies} 

To examine if our SLC hybrid functionals also improve upon the other functionals for core excitation energies, we take the 38 core excitation energies of 13 molecules from Ref.\ \cite{src}, containing 
a total of 15 core$\to$valence and 23 core$\to$Rydberg excitation energies for the first- and second-row nuclei (from the $1s$ core orbitals). In conventional TDDFT, the calculations of core excited 
states can be prohibitively expensive, owing to the large number of roots required to obtain the high energy core excited states. Following Besley {\it et al.} \cite{src}, we perform TDDFT calculations 
using the Tamm-Dancoff approximation (TDA) \cite{ex} within a reduced single excitation space (which includes only excitations from the core orbitals of interest) \cite{redu}, to reduce the 
computational costs of core excitation energies. The calculations are performed with the 6-311(2+,2+)G** basis set and EML(100,302) grid. 

For the core excitation energies (see \Cref{table:ce}), PBE, LC-$\omega$PBE, and $\omega$B97 perform very poorly, while $\omega$B97X, $\omega$B97X-D, and $\omega$B97X-D3 only have 
minor improvement, due to the small fraction of HF exchange in the SR region. By contrast, SLC-PBE-D3 and SLC-B97-D3 perform comparably, slightly improve upon SLC-LDA-D3, and 
significantly outperform PBE, LC-$\omega$PBE, the $\omega$B97 series, and perhaps, other LC hybrid functionals adopting a small fraction of HF exchange in the SR region. For the core 
excitation energies, the statistical errors associated with SLC-PBE-D3 and SLC-B97-D3 are about one order of magnitude smaller than those associated with PBE, LC-$\omega$PBE, and the 
$\omega$B97 series! Therefore, the inclusion of a very large fraction of HF exchange at small interelectronic distances is also important for accurately describing core excitation energies. 
While we do not consider the relativistic corrections here, our comments remain similar for the core excitation energies with the relativistic corrections \cite{supp}.

\subsection{Noncovalent interactions} 

For the noncovalent interactions of the S66 set \cite{S66}, the performance of the functionals is evaluated using the 6-311++G(3df,3pd) basis set and EML(99,590) grid, and the counterpoise 
correction \cite{CP} is adopted to reduce the BSSE. As shown in \Cref{table:s66}, PBE and LC-$\omega$PBE perform very poorly for the noncovalent interactions of the S66 set, due to the lack 
of a proper description of middle- and long-range dynamical correlation effects, while all the dispersion-corrected functionals perform reasonably well.

\subsection{Dissociation of symmetric radical cations} 

Due to the pronounced SIEs associated with semilocal density functionals, unphysical fractional charge dissociation can happen, especially for symmetric charged radicals \cite{X2+}. Here, the 
dissociation energy curves of H$_2^+$, He$_2^+$, Ne$_2^+$, and Ar$_2^+$ are calculated using the 6-311++G(3df,3pd) basis set and EML(75,302) grid to examine the performance of the 
functionals upon the SIE problems. The results are compared with the H$_2^+$ curve calculated using the HF theory (exact for any one-electron system) and the He$_2^+$, Ne$_2^+$, and 
Ar$_2^+$ curves calculated using the highly accurate CCSD(T) theory (coupled-cluster theory with iterative singles and doubles and perturbative treatment of triple substitutions) \cite{CCSD(T)}. 

As shown in \Cref{fig:h2p,fig:he2p,fig:ne2p,fig:ar2p}, unphysical barriers indeed appear in the PBE dissociation curves, owing to the significant SIEs of PBE. By contrast, the LC and SLC hybrid 
functionals greatly reduce (or even remove) the unphysical barriers of the dissociation curves, due to the inclusion of 100\% LR-HF exchange. SLC-LDA-D3, adopting the largest 
$\omega_{\text{LR}}$ (0.45 $\text{bohr}^{-1}$), performs best, followed by SLC-PBE-D3, SLC-B97-D3, LC-$\omega$PBE, and $\omega$B97, adopting the second largest 
$\omega_{\text{LR}}$ (0.40 $\text{bohr}^{-1}$).

\subsection{Atomization energies} 

Recently, we have developed the IP131, EA131, and FG131 databases \cite{p15,p22}, consisting of accurate reference values for the 131 vertical IPs, 131 vertical EAs, and 131 fundamental gaps, 
respectively, of 18 atoms and 113 molecules at their experimental geometries. In addition, we have developed the AE113 database \cite{p22}, which contains accurate reference values for the 
atomization energies of 113 molecules in the IP131 database. Here, we examine the performance of the functionals on the AE113, IP131, EA131, and FG131 databases, 
using the 6-311++G(3df,3pd) basis set and EML(75,302) grid. 

As shown in \Cref{table:ae}, owing to their flexible functional forms, SLC-B97-D3 and the $\omega$B97 series are comparable in performance, more accurate than SLC-PBE-D3, and much more 
accurate than PBE, LC-$\omega$PBE, and SLC-LDA-D3. Interestingly, SLC-PBE-D3 performs better than LC-$\omega$PBE, possibly due to the noticeable deviation of their HF exchange 
operators in the region of 0.5 bohr $\lesssim r_{12} \lesssim$ 0.8 bohr (where the fractions of HF exchange adopted in SLC-PBE-D3, $\omega$B97X-D, and $\omega$B97X-D3 are very similar!).

\subsection{Vertical ionization potentials} 

The vertical IP of a molecule (containing $N$ electrons) is defined as 
\begin{equation}\label{eq:IP1} 
\text{IP}(1) = E_{\text{total}}(N-1) - E_{\text{total}}(N), 
\end{equation} 
where $E_{\text{total}}(N)$ is the total energy of the $N$-electron system. For the exact KS-DFT, the vertical IP of a molecule is the same as the minus HOMO energy of the 
molecule \cite{IP1,IP2,IP3,IP4,IP5,IP6}, 
\begin{equation}\label{eq:IP2} 
\text{IP}(2) = -{\epsilon}_{\text{HOMO}}(N). 
\end{equation} 
However, for an approximate XC density functional in KS-DFT, the computed IP(1) and IP(2) values may be different, showing the accuracy of the predicted total energies and HOMO energies, 
respectively. 

Here, we examine the accuracy of the functionals on the IP131 database \cite{p15}, and summarize our results in \Cref{table:ip}. 
For IP(1), the $\omega$B97 series, SLC-PBE-D3, and SLC-B97-D3 are comparable in performance, outperforming the other functionals. 
For IP(2), LC-$\omega$PBE, $\omega$B97, SLC-PBE-D3, and SLC-B97-D3, which adopt $\omega_{\text{LR}}$ = 0.40 bohr$^{-1}$, perform comparably, and outperform the other functionals. 
By contrast, PBE severely underestimates IP(2), due to the incorrect XC potential asymptote. 
For the IP(1) and IP(2) values, SLC-PBE-D3 and SLC-B97-D3 achieve the best performance, followed closely by $\omega$B97.

\subsection{Vertical electron affinities} 

The vertical EA of a molecule is defined as 
\begin{equation}\label{eq:EA1} 
\text{EA}(1) = E_{\text{total}}(N) - E_{\text{total}}(N+1). 
\end{equation} 
By comparing Eq.\ (\ref{eq:IP1}) with Eq.\ (\ref{eq:EA1}), the vertical EA of a molecule is identical to the vertical IP of the corresponding anion, which is, for the exact KS-DFT, 
the minus HOMO energy of the anion, 
\begin{equation}\label{eq:EA2} 
\text{EA}(2) = -{\epsilon}_{\text{HOMO}}(N+1). 
\end{equation} 
In addition, the vertical EA of a molecule is traditionally approximated by the minus lowest unoccupied molecular orbital (LUMO) energy of the molecule, 
\begin{equation}\label{eq:EA3} 
\text{EA}(3) = -{\epsilon}_{\text{LUMO}}(N). 
\end{equation} 
Nonetheless, even for the exact KS-DFT, there is a fundamental difference between EA(3) and EA(2), owing to the derivative discontinuity $\Delta_{xc}$ \cite{DD1,DD2,DD3,DD4,DD5,IP2,IP6} of 
$E_{xc}[\rho]$: $\text{EA}(3) - \text{EA}(2) = {\epsilon}_{\text{HOMO}}(N+1) - {\epsilon}_{\text{LUMO}}(N) = \Delta_{xc}$. 
Hybrid density functionals, which belong to the generalized Kohn-Sham (GKS) method \cite{GKS} (not pure KS-DFT), effectively capture a fraction of $\Delta_{xc}$ of $E_{xc}[\rho]$ in KS-DFT. 
A recent study has found that the difference between ${\epsilon}_{\text{HOMO}}(N+1)$ and ${\epsilon}_{\text{LUMO}}(N)$ is small for LC hybrid functionals \cite{EA}. Therefore, EA(3) is expected 
to be close to EA(2) (i.e., the true vertical EA) for LC hybrid functionals. 

Here, the accuracy of the functionals on the EA131 database \cite{p15,p22} is investigated. 
As shown in \Cref{table:ea}, all the functionals perform comparably for EA(1). However, for the EA(2) and EA(3) values, the LC and SLC hybrid functionals perform much better than PBE, showing 
the importance of LR-HF exchange in frontier orbital energies. Note that PBE significantly underestimates EA(2), due to the incorrect asymptotic behavior of the XC potential. However, for EA(3), 
there is a fortuitous cancellation of errors in the vertical EA calculated using PBE, as ${\epsilon}_{\text{LUMO}}(N)$ is incorrectly upshifted due to the incorrect PBE XC potential asymptote, 
effectively capturing a fraction of $\Delta_{xc}$ \cite{p22}. Similar to the LC hybrid functionals, $\Delta_{xc}$ is also found to be close to zero for the SLC hybrid functionals, which can be attributed 
to the LR-HF exchange adopted in the LC and SLC hybrid functionals. 
For the EA(1), EA(2), and EA(3) values, SLC-PBE-D3 performs best, followed by SLC-B97-D3.

\subsection{Fundamental gaps} 

The fundamental gap $E_{g}$ of a molecule is the difference between the vertical IP and EA of the molecule, i.e., $E_{g} = \text{IP} - \text{EA}$. 
As mentioned previously, there are various ways of calculating the vertical IP and EA in KS-DFT. Here, we adopt the following three popular ways to calculate $E_{g}$: 
\begin{align} 
& E_{g}(1) = \text{IP}(1) - \text{EA}(1) = E_{\text{total}}(N-1) + E_{\text{total}}(N+1) - 2E_{\text{total}}(N) \label{eq:Eg1}\\ 
& E_{g}(2) = \text{IP}(2) - \text{EA}(2) = {\epsilon}_{\text{HOMO}}(N+1) - {\epsilon}_{\text{HOMO}}(N) \label{eq:Eg2}\\ 
& E_{g}(3) = \text{IP}(2) - \text{EA}(3) = {\epsilon}_{\text{LUMO}}(N) - {\epsilon}_{\text{HOMO}}(N) \label{eq:Eg3} 
\end{align} 
Note that $E_{g}(3)$ is the HOMO-LUMO gap in KS-DFT (i.e., the KS gap). For the exact KS-DFT, both $E_{g}(1)$ and $E_{g}(2)$ lead to the exact fundamental gap, but there is a distinct 
difference between $E_{g}(2)$ and $E_{g}(3)$ (i.e., the energy-gap problem), due to the $\Delta_{xc}$ of $E_{xc}[\rho]$: $E_{g}(2) - E_{g}(3) = \text{EA}(3) - \text{EA}(2) = \Delta_{xc}$. 
For the LC and SLC hybrid functionals, as EA(3) is close to EA(2), $E_{g}(3)$ should be close to $E_{g}(2)$ (i.e., the true fundamental gap). 

Here, we assess the accuracy of the functionals on the FG131 database \cite{p15,p22}. 
As shown in \Cref{table:fg}, for $E_{g}(1)$, the $\omega$B97 series, SLC-PBE-D3, and SLC-B97-D3 are comparable in performance, outperforming the other functionals. 
For $E_{g}(2)$, SLC-PBE-D3 and SLC-B97-D3 perform best, followed closely by $\omega$B97. 
For $E_{g}(3)$, PBE performs worst due to the lack of $\Delta_{xc}$, while SLC-PBE-D3, SLC-B97-D3, and $\omega$B97 perform well for the energy-gap problems here. 
For the $E_{g}(1)$, $E_{g}(2)$, and $E_{g}(3)$ values, SLC-PBE-D3 ranks first, SLC-B97-D3 ranks second, and $\omega$B97 ranks third.

\subsection{Valence and Rydberg excitation energies} 

To examine the performance of the functionals on valence and Rydberg excitation energies, TDDFT calculations are performed on five molecules, involving nitrogen gas (N$_2$), carbon monoxide 
(CO), water (H$_2$O), ethylene (C$_2$H$_4$), and formaldehyde (CH$_2$O), using the 6-311(2+,2+)G** basis set and EML(99,590) grid. The experimental excitation energies are taken from 
Ref.\ \cite{ex}. 

As shown in \Cref{table:vre}, all the functionals perform reasonably well for the valence excitation energies. However, PBE severely underestimates the Rydberg excitation energies due to the 
incorrect XC potential asymptote, while the LC and SLC hybrid functionals perform reasonably well here.

\subsection{Long-range charge-transfer excitation energies} 

Dreuw {\it et al.} have shown that the correct CT excitation energy from the HOMO of a donor to the LUMO of an acceptor should possess the following asymptote \cite{R15}: 
\begin{equation} 
\omega_{\text{CT}}(R \to \infty) \approx \text{IP}_{\text{D}} - \text{EA}_{\text{A}} - 1/R, 
\end{equation} 
where $\text{IP}_{\text{D}}$ is the IP of the donor, $\text{EA}_{\text{A}}$ is the EA of the acceptor, and $R$ is the intermolecular distance. 

Following Dreuw {\it et al.}, we perform TDDFT calculations for the lowest CT excitation energy between ethylene and tetrafluoroethylene with a separation of $R$, using the 6-31G* basis set and 
EML(99,590) grid. High-level {\it ab initio} results obtained with the symmetry-adapted-cluster configuration-interaction (SAC-CI) method are taken from Tawada {\it et al.} for comparison \cite{CT_ref}. 
Unsurprisingly, the LC and SLC hybrid functionals, which retain 100\% LR-HF exchange, yield the correct $(-1/R)$ asymptote in the calculated $\omega_{\text{CT}}(R)$ (see \Cref{fig:ct2}). 
Nevertheless, as shown in \Cref{fig:ct1}, the long-range CT excitation energies are rather sensitive to the LR behavior of the HF exchange operator (i.e., $\omega_{\text{LR}}$), and relatively 
insensitive to the SR behavior of the HF exchange operator (i.e., $\omega_{\text{SR}}$). SLC-LDA-D3, which adopts the largest $\omega_{\text{LR}}$ (0.45 $\text{bohr}^{-1}$), performs best, 
followed by SLC-B97-D3, SLC-PBE-D3, LC-$\omega$PBE, and $\omega$B97, which adopt the second largest $\omega_{\text{LR}}$ (0.40 $\text{bohr}^{-1}$).

\section{Conclusions} 

In summary, we have proposed the SLC hybrid scheme employing 100\% HF exchange at both zero and infinite interelectronic distances, wherein three SLC hybrid density functionals with the D3 
dispersion corrections have been developed. Owing to a very large fraction of HF exchange in the SR region, our SLC-LDA-D3, SLC-PBE-D3, and SLC-B97-D3 functionals yield much more accurate 
core ionization and excitation energies than LC-$\omega$PBE and the $\omega$B97 series. Besides, due to a similar fraction of HF exchange in the MR and LR regions, SLC-PBE-D3 and 
SLC-B97-D3 are generally comparable or superior to LC-$\omega$PBE and the $\omega$B97 series, respectively, in performance, for many other test sets, such as dissociation of symmetric radical 
cations, atomization energies, vertical IPs, vertical EAs, fundamental gaps, and valence, Rydberg, and long-range CT excitation energies. For noncovalent interactions, SLC-LDA-D3, SLC-PBE-D3, 
SLC-B97-D3, and the other dispersion-corrected functionals perform reasonably well. Relative to $\omega$B97X-D, SLC-B97-D3 provides significant improvement for core ionization and excitation 
energies and noticeable improvement for the SIE, asymptote, energy-gap, and CT problems, while performing similarly for thermochemistry, kinetics, and noncovalent interactions. 

By construction, the SLC hybrid scheme can perform reasonably well for the properties sensitive to the SR (e.g., core ionization and excitation energies), MR (e.g., thermochemistry and kinetics), and 
LR (e.g., the SIE, asymptote, energy-gap, and CT problems) behavior of the HF exchange operator. For the properties insensitive to the HF exchange operator (e.g., noncovalent interactions), the 
SLC hybrid scheme does not necessarily yield good accuracy. Nevertheless, to provide an accurate description of noncovalent interactions, the SLC hybrid scheme can be combined with the DFT-D 
schemes, the double-hybrid schemes, and fully nonlocal correlation density functionals for van der Waals interactions. Alternatively, the SLC hybrid scheme can also be extended to the recently 
developed MGGAs with medium-range correlation relevant for noncovalent interactions (e.g., the MGGA$\_$MS family \cite{MGGA_MS1,MGGA_MS2,MGGA_MS3} and SCAN \cite{SCAN}), 
provided that the corresponding MGGA exchange functionals of the MR operator (see Eq.\ (\ref{eq:MR-DFA})) are devised.

\begin{acknowledgments} 

This work was supported by the Ministry of Science and Technology of Taiwan (Grant No.\ MOST104-2628-M-002-011-MY3), National Taiwan University (Grant No.\ NTU-CDP-105R7818), 
the Center for Quantum Science and Engineering at NTU (Subproject Nos.:\ NTU-ERP-105R891401 and NTU-ERP-105R891403), and the National Center for Theoretical Sciences of Taiwan. 

\end{acknowledgments} 

\bibliographystyle{jcp}

\newpage 
\begin{figure} 
\includegraphics[scale=0.95]{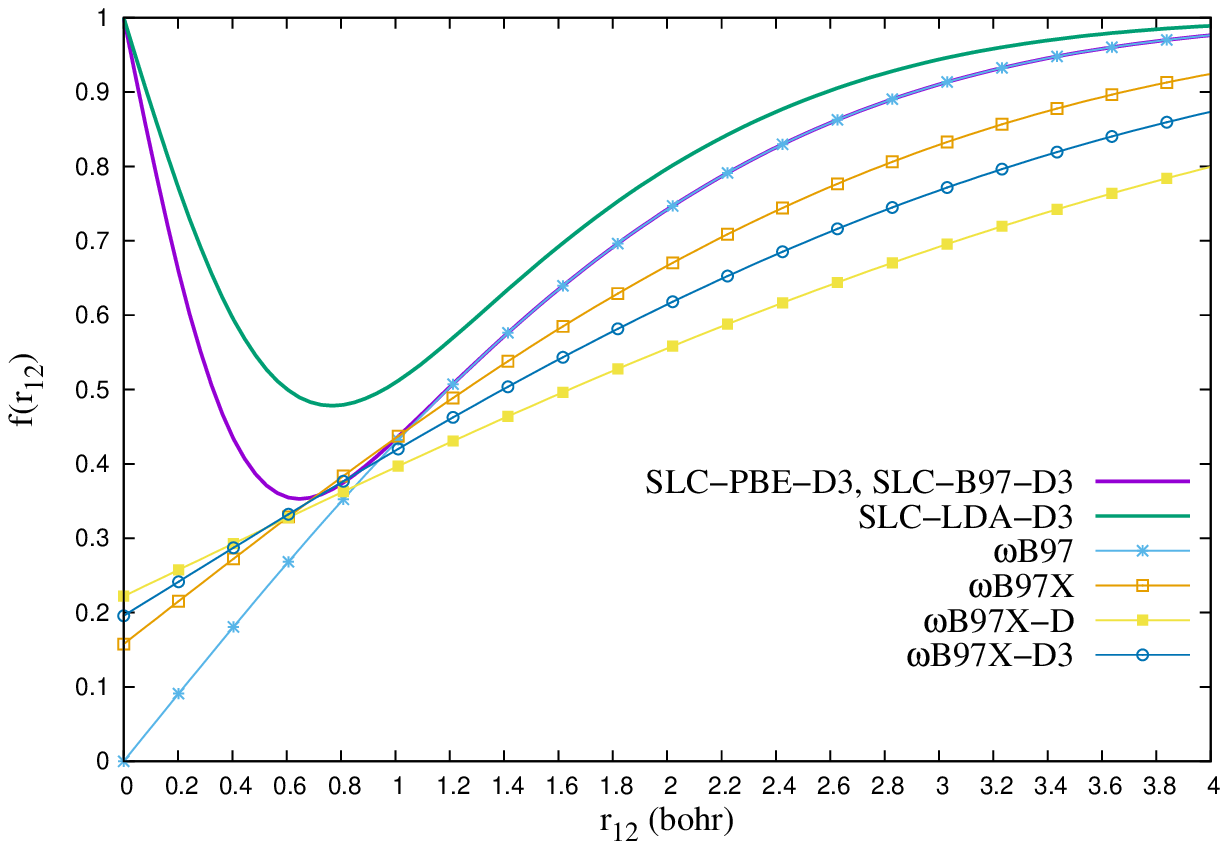} 
\caption{\label{fig:erf} 
Fraction of HF exchange $f(r_{12})$ as a function of the interelectronic distance $r_{12}$, for SLC-LDA-D3, SLC-PBE-D3, SLC-B97-D3, and the $\omega$B97 series.} 
\end{figure} 

\newpage 
\begin{figure} 
\includegraphics[scale=0.95]{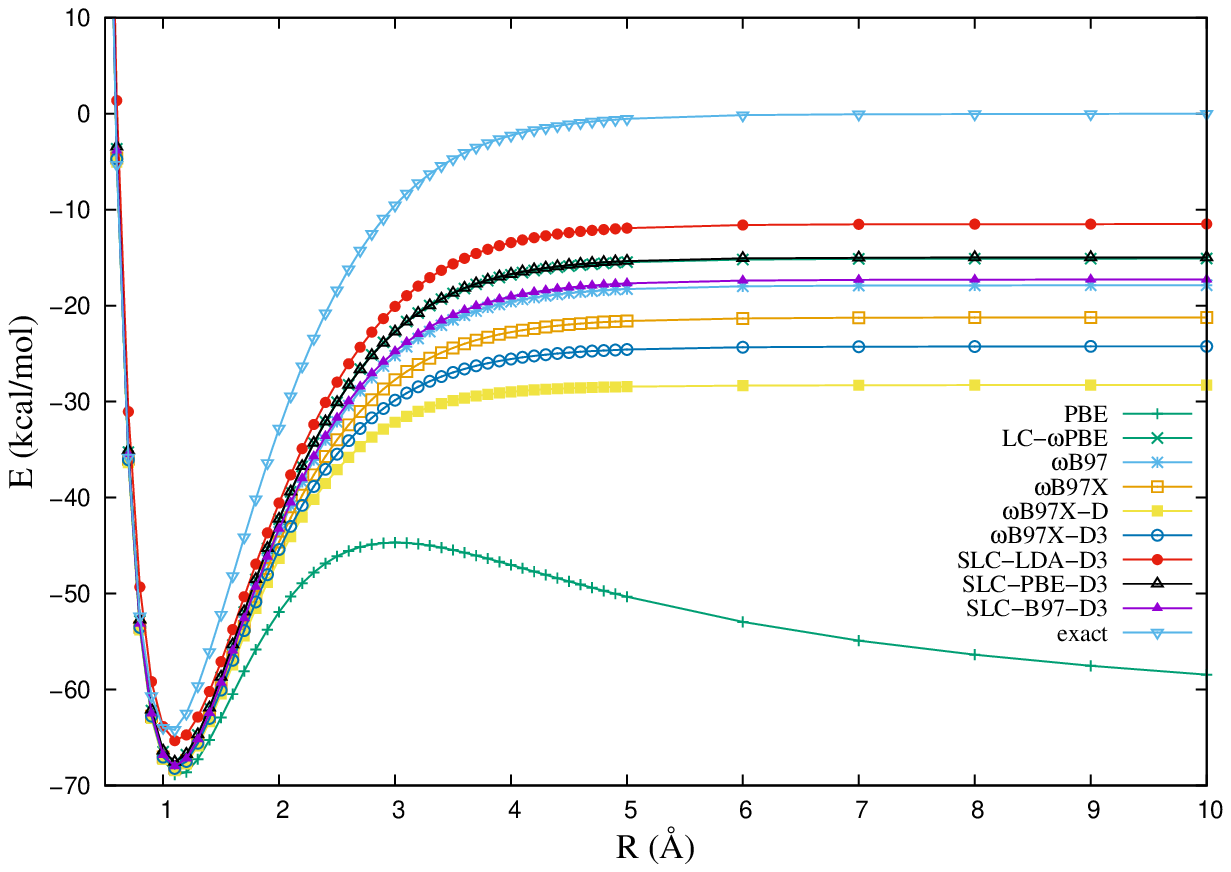} 
\caption{\label{fig:h2p} 
Dissociation energy curve of $\text{H}_{2}^{+}$. Zero level is set to \textit{E}(H) + \textit{E}($\text{H}^{+}$) for each method.} 
\end{figure} 

\newpage 
\begin{figure} 
\includegraphics[scale=0.95]{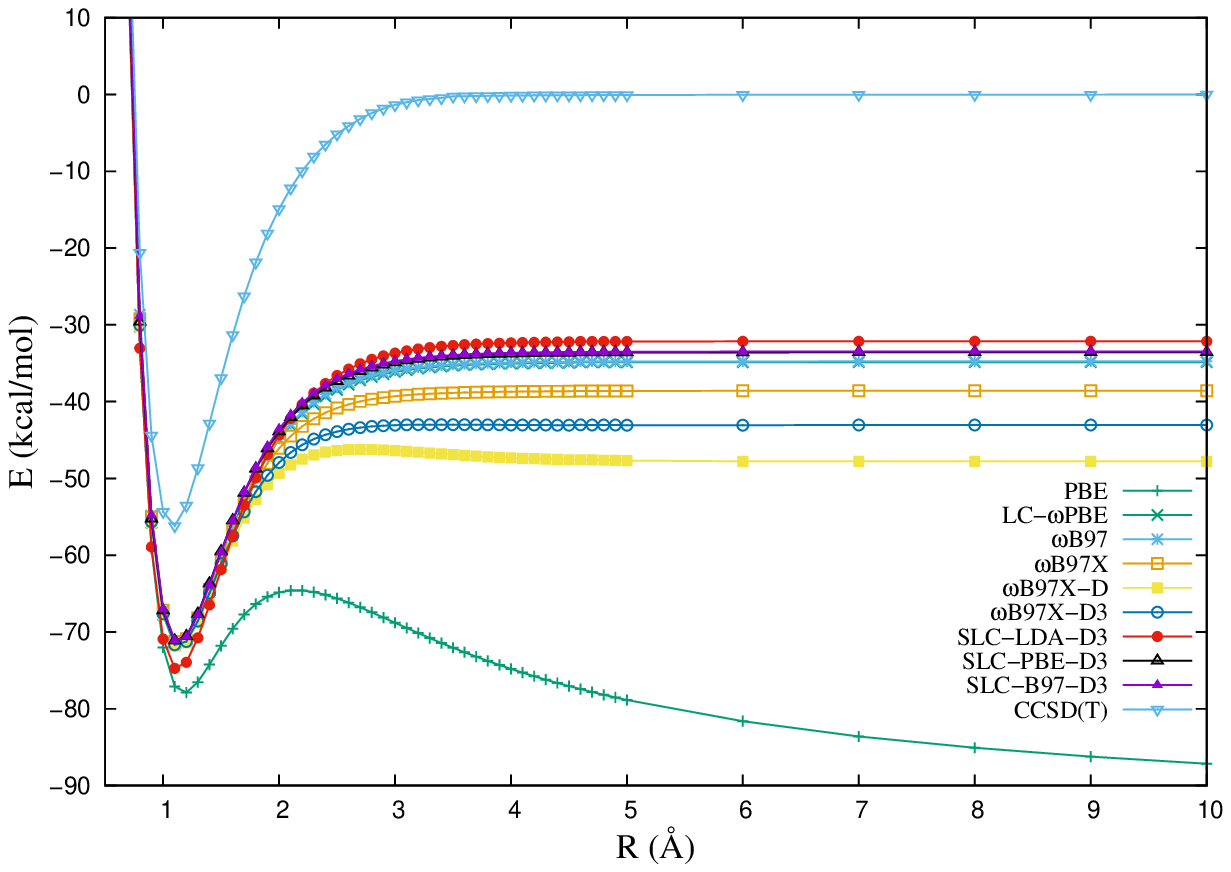} 
\caption{\label{fig:he2p} 
Dissociation energy curve of $\text{He}_{2}^{+}$. Zero level is set to \textit{E}(He) + \textit{E}($\text{He}^{+}$) for each method.} 
\end{figure} 

\newpage 
\begin{figure} 
\includegraphics[scale=0.95]{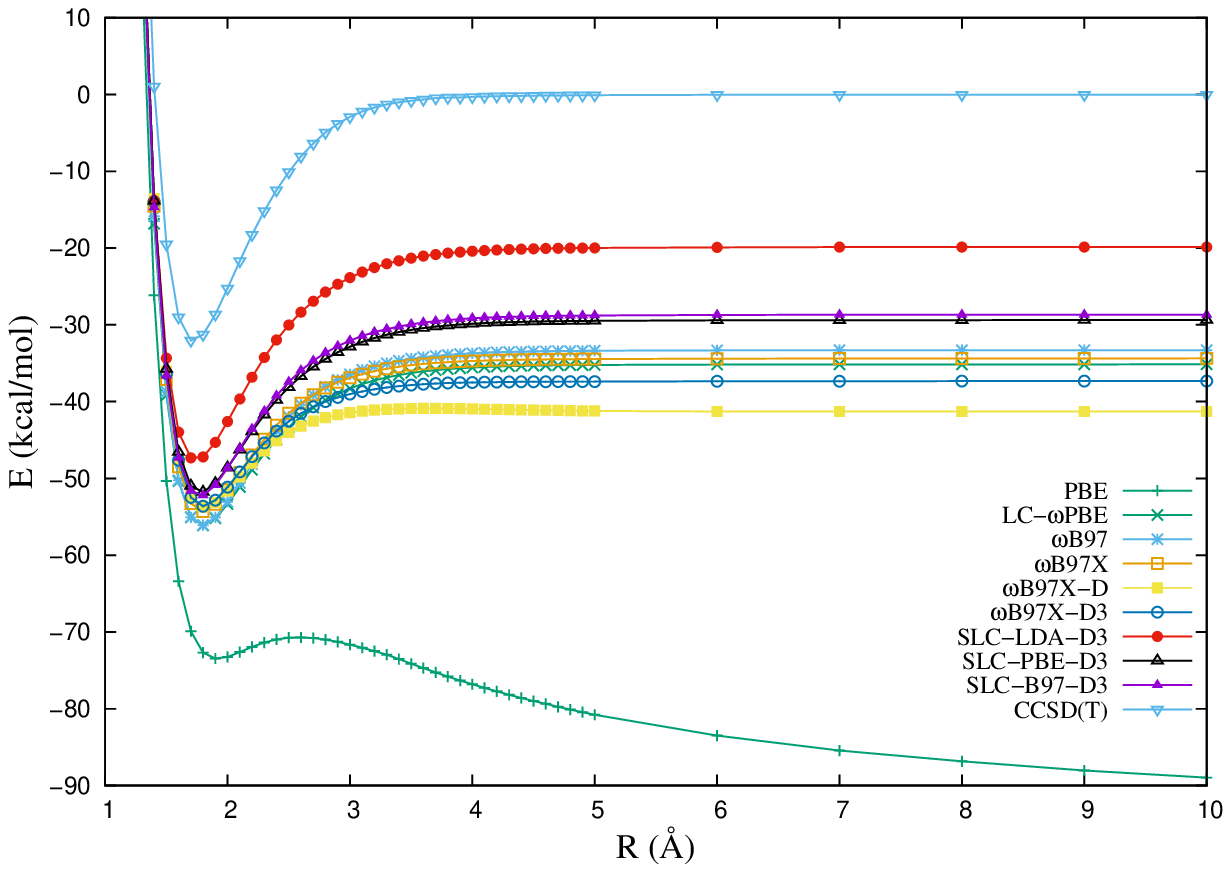} 
\caption{\label{fig:ne2p} 
Dissociation energy curve of $\text{Ne}_{2}^{+}$. Zero level is set to \textit{E}(Ne) + \textit{E}($\text{Ne}^{+}$) for each method.} 
\end{figure} 

\newpage 
\begin{figure} 
\includegraphics[scale=0.95]{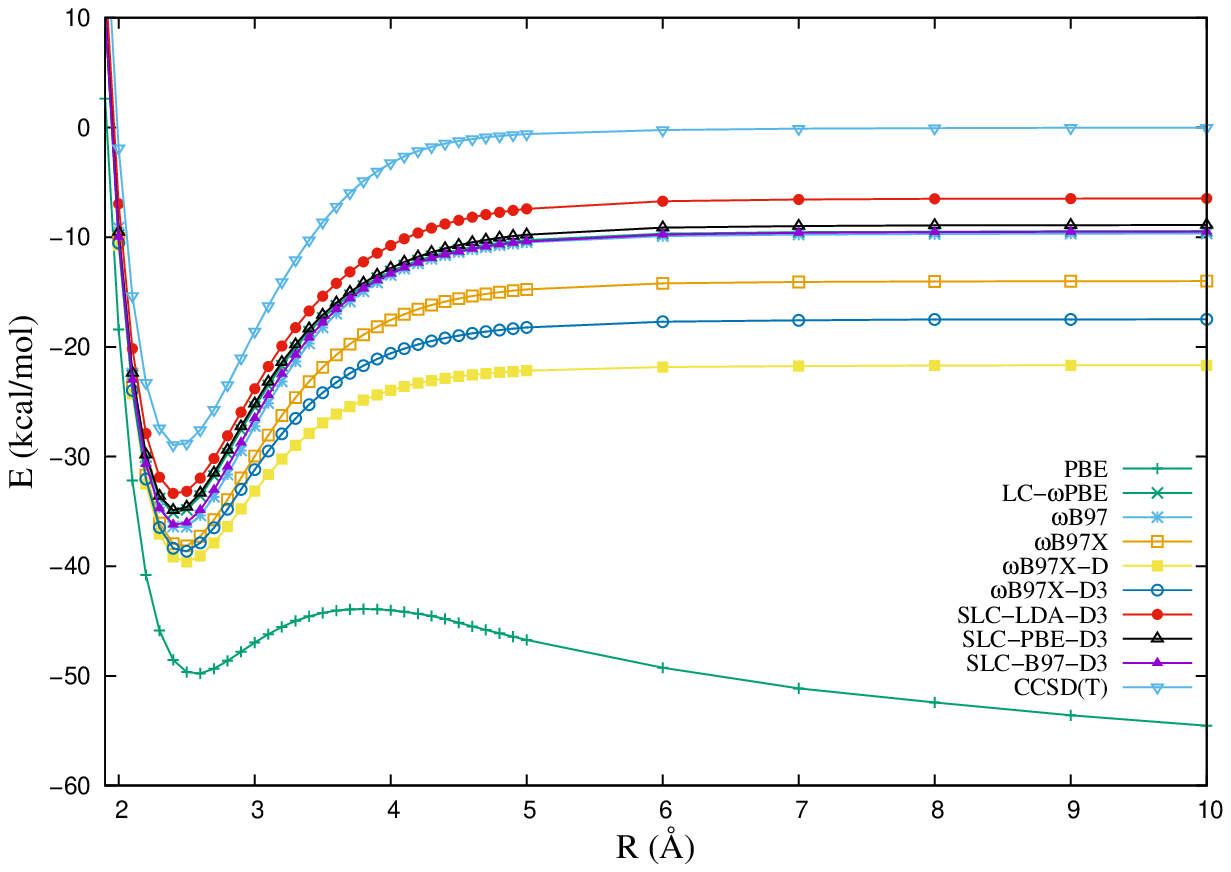} 
\caption{\label{fig:ar2p} 
Dissociation energy curve of $\text{Ar}_{2}^{+}$. Zero level is set to \textit{E}(Ar) + \textit{E}($\text{Ar}^{+}$) for each method.} 
\end{figure} 

\newpage 
\begin{figure} 
\includegraphics[scale=0.95]{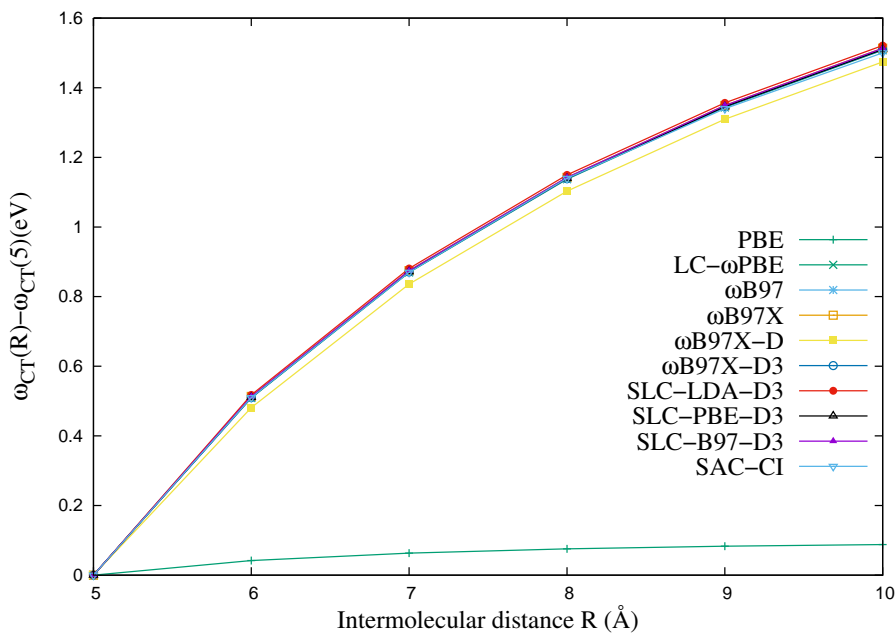} 
\caption{\label{fig:ct2} 
Relative excitation energy for the lowest CT excitation of C$_2$H$_4$$\cdots$C$_2$F$_4$ dimer along the intermolecular distance $R$ (in $\text{\AA}$). 
The excitation energy at 5 $\text{\AA}$ is set to zero for each method.} 
\end{figure} 

\newpage 
\begin{figure} 
\includegraphics[scale=0.95]{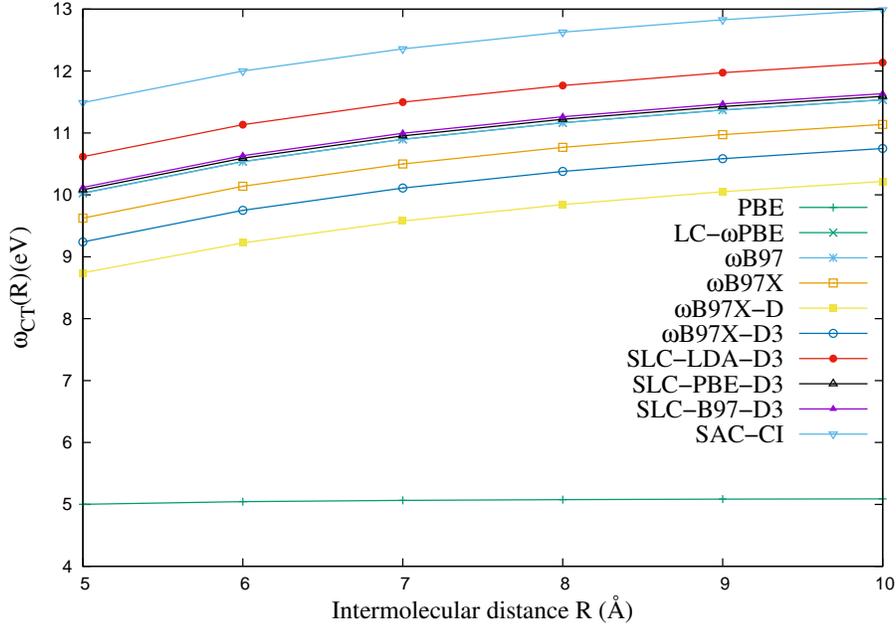} 
\caption{\label{fig:ct1} 
The lowest CT excitation energy of C$_2$H$_4$$\cdots$C$_2$F$_4$ dimer along the intermolecular distance $R$ (in $\text{\AA}$).} 
\end{figure} 

\newpage 
\begin{table*} 
\caption{\label{table:para} 
Optimized parameters for SLC-LDA-D3, SLC-PBE-D3, and SLC-B97-D3. Here, $\omega_{\text{SR}}$ and $\omega_{\text{LR}}$ are defined 
in \Cref{eq:SLR-HF,eq:MR-LDA,eq:MR-PBE,eq:MR-B97}, $s_{r,6}$ and $s_{r,8}$ are defined in \Cref{eq:d3}, and 
the others are defined in \Cref{eq:MR-B97,eq:b97css,eq:b97cab}.} 
\begin{ruledtabular} 
\begin{tabular}{lrrr} 
                                   & SLC-B97-D3 & SLC-PBE-D3 & SLC-LDA-D3 \\ 
\hline 
$\omega_{\text{SR}}\ (\text{bohr}^{-1})$  & 2.0    &  2.0       &  1.5      \\
$\omega_{\text{LR}}\ (\text{bohr}^{-1})$  & 0.40  &  0.40     &  0.45    \\
$s_{r,6}$                         & 1.298          &  1.179   &  1.129  \\
$s_{r,8}$                         & 1.277          &  1.123   &  1.131  \\
$c_{x\sigma,0}$              & 1.000000    &              &             \\
$c_{x\sigma,1}$              & 1.469313    &              &             \\
$c_{x\sigma,2}$              & -6.185202   &              &             \\
$c_{x\sigma,3}$              & 23.053635   &             &             \\
$c_{x\sigma,4}$              & -16.353923  &             &             \\
$c_{c\sigma\sigma,0}$    &  1.000000    &             &             \\
$c_{c\sigma\sigma,1}$    & -2.154721    &             &             \\
$c_{c\sigma\sigma,2}$    & 10.271378   &             &             \\
$c_{c\sigma\sigma,3}$    & -23.966521  &             &             \\
$c_{c\sigma\sigma,4}$    & 15.345722   &             &             \\
$c_{c\alpha\beta,0}$        & 1.000000     &            &             \\
$c_{c\alpha\beta,1}$        & 4.460711     &            &             \\
$c_{c\alpha\beta,2}$        & -25.043202  &            &              \\
$c_{c\alpha\beta,3}$        & 22.506558   &            &              \\
$c_{c\alpha\beta,4}$        & -4.114590    &            &              \\
\end{tabular} 
\end{ruledtabular}  
\end{table*} 

\newpage 
\begin{table*} 
\scriptsize 
\caption{\label{table:train} 
Statistical errors (in kcal/mol) of the training set. PBE, LC-$\omega$PBE, and LC-$\omega$PBE-D3 (statistical errors given in parentheses) 
were not particularly parametrized using this training set.} 
\begin{ruledtabular} 
\begin{tabular}{lrrrrrrrrrr} 
System & Error & PBE   & LC-$\omega$PBE(-D3) & $\omega$B97 & $\omega$B97X & $\omega$B97X-D & $\omega$B97X-D3 & SLC-LDA-D3 & SLC-PBE-D3 & SLC-B97-D3 \\ 
\hline 
G3/99  & MSE   & 20.90 & 3.12 (4.79)           & -0.29       & -0.20        & -0.24          & -0.14           & 2.63       & -0.57      & -0.32      \\
(223)   & MAE   & 21.51 & 5.86 (7.17)          & 2.63        & 2.13         & 1.93           & 2.06            & 8.84       & 4.49       & 2.63       \\
             & rms   & 26.30 & 7.43 (9.02)          & 3.58        & 2.88         & 2.77           & 2.81            & 11.09      & 5.91       & 3.49       \\
IP     & MSE   & 0.04  & 2.86 (2.85)          & -0.50       & -0.14        & 0.20           & 0.07            & 11.60      & 1.85       & -0.22      \\
(40)   & MAE   & 3.44  & 4.29 (4.29)          & 2.68        & 2.69         & 2.75           & 2.66            & 11.60      & 3.74       & 2.54       \\
	       & rms   & 4.35  & 5.39 (5.39)          & 3.60        & 3.59         & 3.62           & 3.53            & 12.40      & 4.70       & 3.45       \\
EA     & MSE   & 1.72  & 0.18 (0.18)          & 1.52        & -0.47        & 0.07           & -0.37           & 8.96       & -0.54      & -1.66      \\
(25)   & MAE   & 2.42  & 3.00 (3.01)          & 2.72        & 2.04         & 1.91           & 1.93            & 8.96       & 3.05       & 2.66       \\
           & rms   & 3.06  & 3.50 (3.51)          & 3.11        & 2.57         & 2.38           & 2.41            & 9.62       & 3.52       & 3.06       \\
PA     & MSE   & -0.83 & 0.86 (0.94)          & 0.67        & 0.56         & 1.42           & 1.10            & -1.91      & 0.84       & 0.80       \\
(8)    & MAE   & 1.60  & 1.41 (1.45)          & 1.48        & 1.21         & 1.50           & 1.29            & 2.31       & 1.36       & 1.44       \\
           & rms   & 1.91  & 2.04 (2.08)          & 2.18        & 1.70         & 2.05           & 1.92            & 2.54       & 2.01       & 2.19       \\
NHTBH  & MSE   & -8.52 & 1.39 (1.01)          & 1.32        & 0.55         & -0.45          & 0.04            & 1.99       & 1.29       & 1.38       \\
(38)   & MAE   & 8.62  & 2.47 (2.28)          & 2.32        & 1.75         & 1.51           & 1.53            & 3.32       & 2.38       & 2.13       \\
           & rms   & 10.61 & 3.07 (2.83)          & 2.82        & 2.08         & 2.00           & 1.89            & 3.77       & 2.86       & 2.55       \\
HTBH   & MSE   & -9.67 & -0.77 (-1.23)         & -0.66       & -1.55        & 2.57           & -2.08           & -0.27      & -1.03      & -0.96      \\
(38)   & MAE   & 9.67  & 1.39 (1.59)          & 2.11        & 2.27         & 2.70           & 2.40            & 1.99       & 1.41       & 2.04       \\
           & rms   & 10.37 & 1.90 (2.07)          & 2.47        & 2.60         & 3.10           & 2.75            & 2.59       & 1.77       & 2.33       \\
S22    & MSE   & 2.71  & 2.82 (-0.08)          & 0.10        & 0.47         & -0.14          & -0.07           & 0.34       & 0.11       & -0.20      \\
(22)   & MAE   & 2.71  & 2.82 (0.26)          & 0.53        & 0.79         & 0.19           & 0.18            & 0.45       & 0.30       & 0.23       \\
           & rms   & 3.73  & 3.58 (0.35)          & 0.63        & 1.11         & 0.25           & 0.25            & 0.61       & 0.39       & 0.33       \\
Total  & MSE   & 10.32 & 2.30 (3.01)          & -0.23       & -0.21        & -0.38          & -0.28           & 3.38       & -0.12      & -0.26      \\
(394)  & MAE   & 14.63 & 4.50 (5.10)          & 2.42        & 2.06         & 1.94           & 1.97            & 7.33       & 3.52       & 2.36       \\
           & rms   & 20.40 & 6.09 (7.15)          & 3.27        & 2.76         & 2.73           & 2.69            & 9.65       & 4.90       & 3.15       \\
\end{tabular} 
\end{ruledtabular} 
\end{table*} 

\newpage 
\begin{table*} 
\scriptsize 
\caption{\label{table:ci} 
Statistical errors (in eV) of the 23 core ionization energies of 14 molecules taken from Ref.\ \cite{CI}. 
The relativistic corrections are not considered.} 
\begin{ruledtabular} 
\begin{tabular}{lrrrrrrrrrr} 
System     & Error & PBE    & LC-$\omega$PBE & $\omega$B97 & $\omega$B97X & $\omega$B97X-D & $\omega$B97X-D3 & SLC-LDA-D3 & SLC-PBE-D3 & SLC-B97-D3 \\ 
\hline 
Core       & MSE   & -26.25 & -20.15         & -19.39      & -14.88       & -13.74         & -14.10          & 4.25       & -2.36      & -1.53      \\
Ionization & MAE   & 26.25  & 20.15          & 19.39       & 14.88        & 13.74          & 14.10           & 4.27       & 2.77       & 2.53       \\
(23)       & rms   & 26.48  & 20.47          & 19.70       & 15.10        & 13.91          & 14.29           & 5.27       & 3.36       & 2.91       \\
\end{tabular} 
\end{ruledtabular}  
\end{table*} 

\newpage 
\begin{table*} 
\scriptsize 
\caption{\label{table:ce} 
Statistical errors (in eV) of the 38 core excitation energies of 13 molecules taken from Ref.\ \cite{src}. 
The relativistic corrections are not considered.} 
\begin{ruledtabular} 
\begin{tabular}{lrrrrrrrrrr} 
State             & Error & PBE   & LC-$\omega$PBE & $\omega$B97 & $\omega$B97X & $\omega$B97X-D & $\omega$B97X-D3 & SLC-LDA-D3 & SLC-PBE-D3 & SLC-B97-D3 \\
\hline 
Core $\to$     & MSE   & -42.32  & -41.30         & -40.31      & -31.95       & -28.74         & -30.05          & 4.81       & -1.46      & -0.38      \\ 
Valence         & MAE   & 42.32   & 41.30          & 40.31       & 31.95        & 28.74          & 30.05           & 5.12       & 2.22       & 2.53       \\ 
(15)               & rms   & 50.91   & 49.93          & 48.96       & 39.29        & 35.56          & 37.08           & 6.25       & 2.91       & 2.84       \\ 
Core $\to$     & MSE   & -32.26  & -29.35         & -28.38      & -22.26       & -20.35         & -21.12          & 3.36       & -2.93      & -1.93      \\ 
Rydberg        & MAE   & 32.26   & 29.35          & 28.38       & 22.26        & 20.35          & 21.12           & 3.50       & 3.22       & 2.94       \\ 
(23)               & rms   & 39.91   & 37.64          & 36.78       & 29.43        & 26.90          & 27.93           & 4.94       & 3.81       & 3.26       \\ 
\end{tabular} 
\end{ruledtabular} 
\end{table*} 

\newpage 
\begin{table*} 
\scriptsize 
\caption{\label{table:s66} 
Statistical errors (in kcal/mol) of the S66 set \cite{S66}.} 
\begin{ruledtabular} 
\begin{tabular}{lrrrrrrrrrr} 
System & Error & PBE  & LC-$\omega$PBE & $\omega$B97 & $\omega$B97X & $\omega$B97X-D & $\omega$B97X-D3 & SLC-LDA-D3 & SLC-PBE-D3 & SLC-B97-D3 \\
\hline
S66    & MSE   & 2.22 & 2.46           & -0.15       & 0.16         & -0.30          & -0.23           & 0.04       & -0.06      & -0.35      \\
(66)   & MAE   & 2.23 & 2.46           & 0.37        & 0.49         & 0.35           & 0.26            & 0.21       & 0.27       & 0.37       \\
           & rms   & 2.75 & 2.80           & 0.47        & 0.65         & 0.51           & 0.35            & 0.30       & 0.35       & 0.46       \\
\end{tabular} 
\end{ruledtabular} 
\end{table*} 

\newpage 
\begin{table*} 
\scriptsize 
\caption{\label{table:ae} 
Statistical errors (in eV) of the AE113 database \cite{p15,p22}.} 
\begin{ruledtabular} 
\begin{tabular}{lrrrrrrrrrr} 
System & Error & PBE  & LC-$\omega$PBE & $\omega$B97 & $\omega$B97X & $\omega$B97X-D & $\omega$B97X-D3 & SLC-LDA-D3 & SLC-PBE-D3 & SLC-B97-D3 \\ 
\hline 
AE113  & MSE   & 0.83 & 0.10           & 0.05        & 0.05         & 0.04           & 0.05            & 0.04       & -0.03      & 0.04       \\
(113)  & MAE   & 0.88 & 0.27           & 0.11        & 0.10         & 0.10           & 0.10            & 0.28       & 0.17       & 0.11       \\
           & rms   & 1.06 & 0.41           & 0.15        & 0.13         & 0.14           & 0.13            & 0.34       & 0.23       & 0.14       \\
\end{tabular} 
\end{ruledtabular} 
\end{table*} 

\newpage 
\begin{table*} 
\scriptsize 
\caption{\label{table:ip} 
Statistical errors (in eV) of the IP131 database \cite{p15}.} 
\begin{ruledtabular} 
\begin{tabular}{lrrrrrrrrrr} 
System     & Error & PBE   & LC-$\omega$PBE & $\omega$B97 & $\omega$B97X & $\omega$B97X-D & $\omega$B97X-D3 & SLC-LDA-D3 & SLC-PBE-D3 & SLC-B97-D3 \\ 
\hline
\multicolumn{11}{l}{$\text{IP}(1) = E_{\text{total}}(N-1) - E_{\text{total}}(N)$} \\
IP131      & MSE   & -0.26 & 0.10           & 0.00        & 0.00         & -0.03          & -0.02           & 0.57       & 0.09       & 0.02       \\
(131)      & MAE   & 0.36  & 0.28           & 0.19        & 0.18         & 0.19           & 0.18            & 0.58       & 0.20       & 0.18       \\
               & rms   & 0.52  & 0.46           & 0.26        & 0.26         & 0.27           & 0.26            & 0.64       & 0.28       & 0.26       \\
\multicolumn{11}{l}{$\text{IP}(2) = -{\epsilon}_{\text{HOMO}}(N)$} \\ 
IP131      & MSE   & -4.40 & -0.15          & -0.24       & -0.48        & -1.01          & -0.71           & 0.61       & -0.09      & -0.18      \\
(131)      & MAE   & 4.40  & 0.42           & 0.40        & 0.51         & 1.01           & 0.72            & 0.70       & 0.36       & 0.37       \\
               & rms   & 4.50  & 0.68           & 0.63        & 0.75         & 1.18           & 0.93            & 0.77       & 0.56       & 0.59       \\
\end{tabular} 
\end{ruledtabular} 
\end{table*} 

\newpage 
\begin{table*} 
\scriptsize 
\caption{\label{table:ea} 
Statistical errors (in eV) of the EA131 database \cite{p15,p22}.} 
\begin{ruledtabular} 
\begin{tabular}{lrrrrrrrrrr} 
System     & Error & PBE   & LC-$\omega$PBE & $\omega$B97 & $\omega$B97X & $\omega$B97X-D & $\omega$B97X-D3 & SLC-LDA-D3 & SLC-PBE-D3 & SLC-B97-D3 \\ 
\hline 
\multicolumn{11}{l}{$\text{EA}(1) = E_{\text{total}}(N) - E_{\text{total}}(N+1)$} \\ 
EA131      & MSE   & 0.10  & -0.11          & -0.24       & -0.19        & -0.14          & -0.16           & 0.23       & -0.08      & -0.22      \\
(131)      & MAE   & 0.21  & 0.36           & 0.34        & 0.31         & 0.26           & 0.29            & 0.33       & 0.27       & 0.32       \\
               & rms   & 0.34  & 0.54           & 0.44        & 0.41         & 0.36           & 0.39            & 0.45       & 0.35       & 0.42       \\
\multicolumn{11}{l}{$\text{EA}(2) = -{\epsilon}_{\text{HOMO}}(N+1)$} \\ 
EA131      & MSE   & -2.03 & 0.00           & -0.13       & -0.17        & -0.32          & -0.23           & 0.40       & 0.08       & -0.10      \\
(131)      & MAE   & 2.03  & 0.34           & 0.34        & 0.32         & 0.39           & 0.33            & 0.47       & 0.30       & 0.33       \\
               & rms   & 2.30  & 0.43           & 0.43        & 0.41         & 0.51           & 0.44            & 0.62       & 0.38       & 0.41       \\
\multicolumn{11}{l}{$\text{EA}(3) = -{\epsilon}_{\text{LUMO}}(N)$} \\ 
EA131      & MSE   & 2.43  & -0.19          & -0.38       & -0.31        & 0.01           & -0.15           & 0.10       & -0.25      & -0.34      \\
(131)      & MAE   & 2.45  & 0.42           & 0.49        & 0.52         & 0.54           & 0.49            & 0.30       & 0.39       & 0.44       \\
               & rms   & 2.72  & 0.51           & 0.59        & 0.62         & 0.63           & 0.57            & 0.40       & 0.47       & 0.53       \\
\end{tabular} 
\end{ruledtabular} 
\end{table*} 

\newpage 
\begin{table*} 
\scriptsize 
\caption{\label{table:fg} 
Statistical errors (in eV) of the FG131 database \cite{p15,p22}.} 
\begin{ruledtabular} 
\begin{tabular}{lrrrrrrrrrr} 
System     & Error & PBE   & LC-$\omega$PBE & $\omega$B97 & $\omega$B97X & $\omega$B97X-D & $\omega$B97X-D3 & SLC-LDA-D3 & SLC-PBE-D3 & SLC-B97-D3 \\ 
\hline 
\multicolumn{11}{l}{$E_{g}(1) = E_{\text{total}}(N-1) + E_{\text{total}}(N+1) - 2E_{\text{total}}(N)$} \\ 
FG131      & MSE   & -0.46 & 0.10           & 0.13        & 0.08         & -0.01          & 0.03            & 0.23       & 0.05       & 0.13       \\ 
(131)      & MAE   & 0.57  & 0.44           & 0.32        & 0.29         & 0.27           & 0.28            & 0.39       & 0.30       & 0.32       \\ 
               & rms   & 0.76  & 0.74           & 0.42        & 0.40         & 0.38           & 0.39            & 0.47       & 0.39       & 0.41       \\ 
\multicolumn{11}{l}{$E_{g}(2) = {\epsilon}_{\text{HOMO}}(N+1) - {\epsilon}_{\text{HOMO}}(N)$} \\ 
FG131      & MSE   & -2.48 & -0.26          & -0.22       & -0.42        & -0.80          & -0.59           & 0.09       & -0.27      & -0.18      \\ 
(131)      & MAE   & 2.48  & 0.50           & 0.44        & 0.50         & 0.81           & 0.62            & 0.47       & 0.43       & 0.43       \\ 
               & rms   & 2.69  & 0.70           & 0.59        & 0.65         & 0.93           & 0.77            & 0.57       & 0.57       & 0.57       \\ 
\multicolumn{11}{l}{$E_{g}(3) = {\epsilon}_{\text{LUMO}}(N) - {\epsilon}_{\text{HOMO}}(N)$} \\ 
FG131      & MSE   & -6.94 & -0.07          & 0.04        & -0.27        & -1.12          & -0.67           & 0.40       & 0.05       & 0.05       \\ 
(131)      & MAE   & 6.94  & 0.53           & 0.49        & 0.57         & 1.15           & 0.77            & 0.54       & 0.45       & 0.45       \\ 
               & rms   & 7.15  & 0.80           & 0.65        & 0.77         & 1.40           & 1.02            & 0.63       & 0.59       & 0.60       \\ 
\end{tabular} 
\end{ruledtabular} 
\end{table*} 

\newpage 
\begin{table*} 
\scriptsize 
\caption{\label{table:vre} 
Statistical errors (in eV) of the 19 valence and 23 Rydberg excitation energies of five molecules (N$_2$, CO, water, ethylene, and formaldehyde) taken from Ref.\ \cite{ex}.} 
\begin{ruledtabular} 
\begin{tabular}{lrrrrrrrrrr} 
State   & Error & PBE   & LC-$\omega$PBE & $\omega$B97 & $\omega$B97X & $\omega$B97X-D & $\omega$B97X-D3 & SLC-LDA-D3 & SLC-PBE-D3 & SLC-B97-D3 \\ 
\hline 
Valence & MSE   & -0.30 & -0.36          & -0.23       & -0.28        & -0.29          & -0.28           & -0.32      & -0.37      & -0.23      \\ 
(19)    & MAE   & 0.31  & 0.39           & 0.27        & 0.30         & 0.30           & 0.29            & 0.37       & 0.40       & 0.27       \\ 
        & rms   & 0.38  & 0.47           & 0.34        & 0.37         & 0.37           & 0.37            & 0.46       & 0.48       & 0.35       \\ 
Rydberg & MSE   & -1.29 & 0.19           & 0.24        & 0.12         & -0.30          & -0.12           & 0.51       & 0.20       & 0.22       \\ 
(23)    & MAE   & 1.29  & 0.29           & 0.28        & 0.21         & 0.35           & 0.22            & 0.51       & 0.28       & 0.27       \\ 
        & rms   & 1.35  & 0.38           & 0.37        & 0.30         & 0.40           & 0.30            & 0.58       & 0.37       & 0.36       \\ 
\end{tabular} 
\end{ruledtabular} 
\end{table*} 

\end{document}